\documentclass[]{piparticle-final}
\usepackage{graphicx}
\usepackage{amsmath}
\usepackage[latin1]{inputenc}

\usepackage{cite} % To compress citation ranges
\usepackage{epstopdf}

\begin{document}

\volume{7}               % To be inserted by Editor
\articlenumber{070017}   % To be inserted by Editor
\journalyear{2015}       % To be inserted by Editor
\editor{E. Mizraji}   % To be inserted by Editor
\reviewers{J. Lin, Department of Physics, Washington College, Maryland, USA.}  % To be inserted by Editor
\received{18 October 2015}     % To be inserted by Editor
\accepted{10 November 2015}   % To be inserted by Editor
\runningauthor{D. Chialvo \itshape{et al.}}  % To be inserted by Editor
\doi{070017}         % To be inserted by Editor

\title{How we move is universal: Scaling in the average shape of human activity}


\author{
Dante R. Chialvo,\cite{inst1} 
Ana Mar\'ia Gonzalez Torrado,\cite{inst2}
Ewa Gudowska-Nowak,\cite{inst3}\\
Jeremi K. Ochab,\cite{inst4}
Pedro Montoya,\cite{inst2}
Maciej A. Nowak,\cite{inst3,inst4}
Enzo Tagliazucchi\cite{inst5}
 }\pipabstract{Human motor activity is constrained by the rhythmicity of the 24 hours circadian cycle, including the usual 12-15 hours sleep-wake cycle. However, activity fluctuations also appear over a wide range of temporal scales, from days to a few seconds, resulting from the concatenation of a myriad of individual smaller motor events. Furthermore,  individuals present different propensity to wakefulness and thus to motor activity throughout the circadian cycle.  Are activity fluctuations across temporal scales intrinsically different, or is there a universal description encompassing them? Is this description also universal across individuals, considering the aforementioned variability?  Here we establish the presence of universality in motor activity fluctuations based on the empirical study of a month of continuous wristwatch accelerometer recordings. We study the scaling of average fluctuations across temporal scales and determine a universal law characterized by critical exponents $\alpha$, $\tau$ and $1/{ \mu}$. Results are highly reminiscent of the universality described for the average shape of avalanches in systems exhibiting crackling noise. Beyond its theoretical relevance, the present results can be important for developing objective markers of healthy  as well as pathological human motor behavior.
 }
\maketitle

\blfootnote{
\begin{theaffiliation}{99}
\institution{inst1} Consejo Nacional de Investigaciones Cient\'ificas y Tecnol\'ogicas (CONICET), Rivadavia 1917, Buenos Aires, Argentina.  \\
\institution{inst2} Institut Universitari d{'}Investigacions en Ci\`encies de la Salut (IUNICS) $\&$  Universitat de les Illes Balears (UIB), Palma de Mallorca, Spain.\\  
\institution{inst3} M. Kac Complex Systems Research Center and M. Smoluchowski Institute of Physics, Jagiellonian University, Krak\'{o}w, Poland. \\
\institution{inst4} Biocomplexity Department, Ma\l{}opolska Center of Biotechnology, Jagiellonian University, Krak\'{o}w, Poland.
\institution{inst5}Institute for Medical Psychology, Christian Albrechts University,  Kiel, Germany.
\end{theaffiliation}
}

\section{Introduction}

The most obvious periodicity of human (as well as animal) motor activity is the circadian twenty four hours modulation. However, smaller fluctuations are evident on a wide range of temporal scales, from days to a few seconds. Data shows that the activity evolves in bursts of all sizes and durations which are known to be scale-invariant \cite {nakamura1, nakamura2, hu2004, amaral, Celia, Christensen, Proekt, krakow1} regardless of the origins and intended consequences of such activity.   Despite the variety of results, the mechanisms underlying the scale-invariant behavior of motor activity remain to be elucidated. Considering the intermittent nature of human motor activity - comprising brief activity excursions separated by periods of quiescence -  a natural approach would be to study the average shape of the events, following recent results \cite{laurson, papaniko, sethna,neuro1} which show that for a large class of processes,  the average shape is a scaling function determined mostly by the temporal correlations of the process and its nonlinearities \cite{baldassarri}. 
 
In the present work, long time series of human motor activity are analyzed, recorded via wristwatch accelerometer, lasting approximately one month. We establish first the presence of truncated scale-invariance in the distribution of the durations of the events as well as in its power spectral density, as described previously in similar type of data. Afterwards, we uncover the average shape of the bursts of activity and derive the scaling function and its associated exponents. Finally, we discuss the origins of such scaling and some possible applications.
 
%% Figure 1%%%%%%%%%%%%%%%%%%%%%%
 \begin{figure}[h]
\centering
\includegraphics[width=.49 \textwidth]{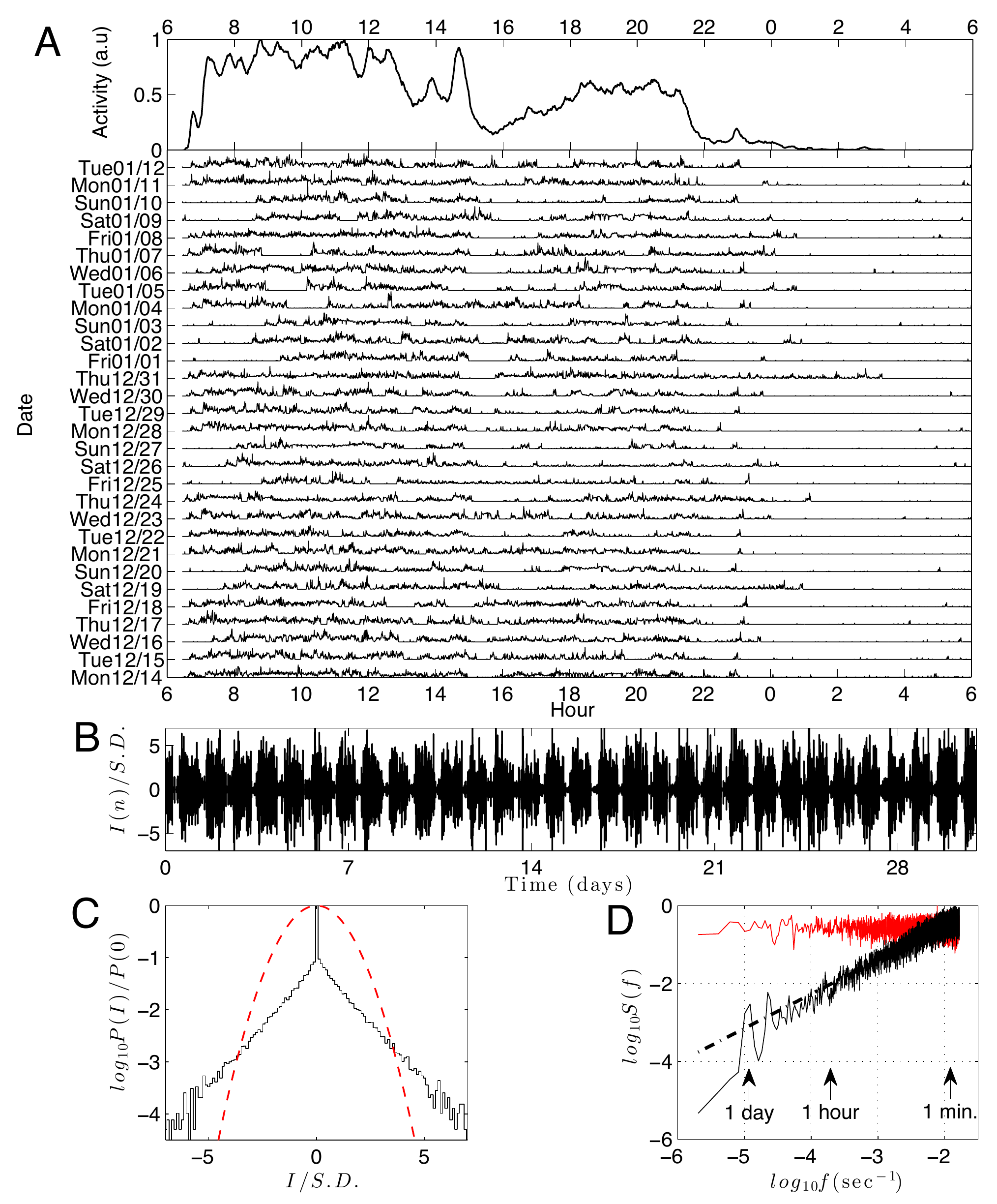}
\caption{Example data set, distribution of successive increments and their spectral power. Panel A: Time series of activity $x(n)$ recorded continuously from a subject during a month. Individual traces correspond to consecutive days. The top subpanel depicts  daily activity averaged over the entire month. Panel B: Time series of successive increments $I(n)= x(n+1)-x(n)$ (normalized by its SD) for the same data. Panel C: Probability density distribution of the time series of successive increments $I(n)$ (continuous line), exhibiting exponential tails (compare with the dotted line, a Gaussian of the same variance).  Panel D: Power spectral density (black line) of the time series of successive increments $I(n)$ of panel B. This is scale invariant  $S(f) \sim f^ \gamma$ with $\gamma= 0.9$ (dashed line). In contrast, for the randomly shuffled increments, the serial correlations vanish and a flat spectral density is obtained (red). }
\end{figure}
%%%%%%%%%%%%%%%%%%%

\section{Materials and methods}

The recordings analyzed were part of a larger study and included six healthy, non-smokers, drug-free volunteers (mean age 50.1 years, S.D. = 6.8). The study was approved by the Bioethics Commission of the University of Isles Baleares (Spain). Participants were informed about the procedures and goals of the study, and provided their written consent. After determining their handedness, each subject was provided with a wristwatch-sized activity recorder (Actiwatch from Mini- Mitter Co., OR, USA) measuring acceleration changes in the forearm in any plane. Each data point of activity corresponded to the number of zero crossings in acceleration larger than 0.01 G  (sampled at 32 Hz and integrated over a 30-second window length).  Records of several thousands of data points were kept in the device's internal memory until being downloaded to a personal computer every week. Subjects wore the device in their non-dominant arm continuously for up to several weeks (mean 28.1 days, S.D.= 4.). After careful visual inspection of the data to exclude sets with gaps (due to subject non-compliance), a combined total of 280 days of data was available for further analysis. 
%%%%%%%%%%Figure 2
\begin{figure}[h]
\centering
\includegraphics[width=.45 \textwidth]{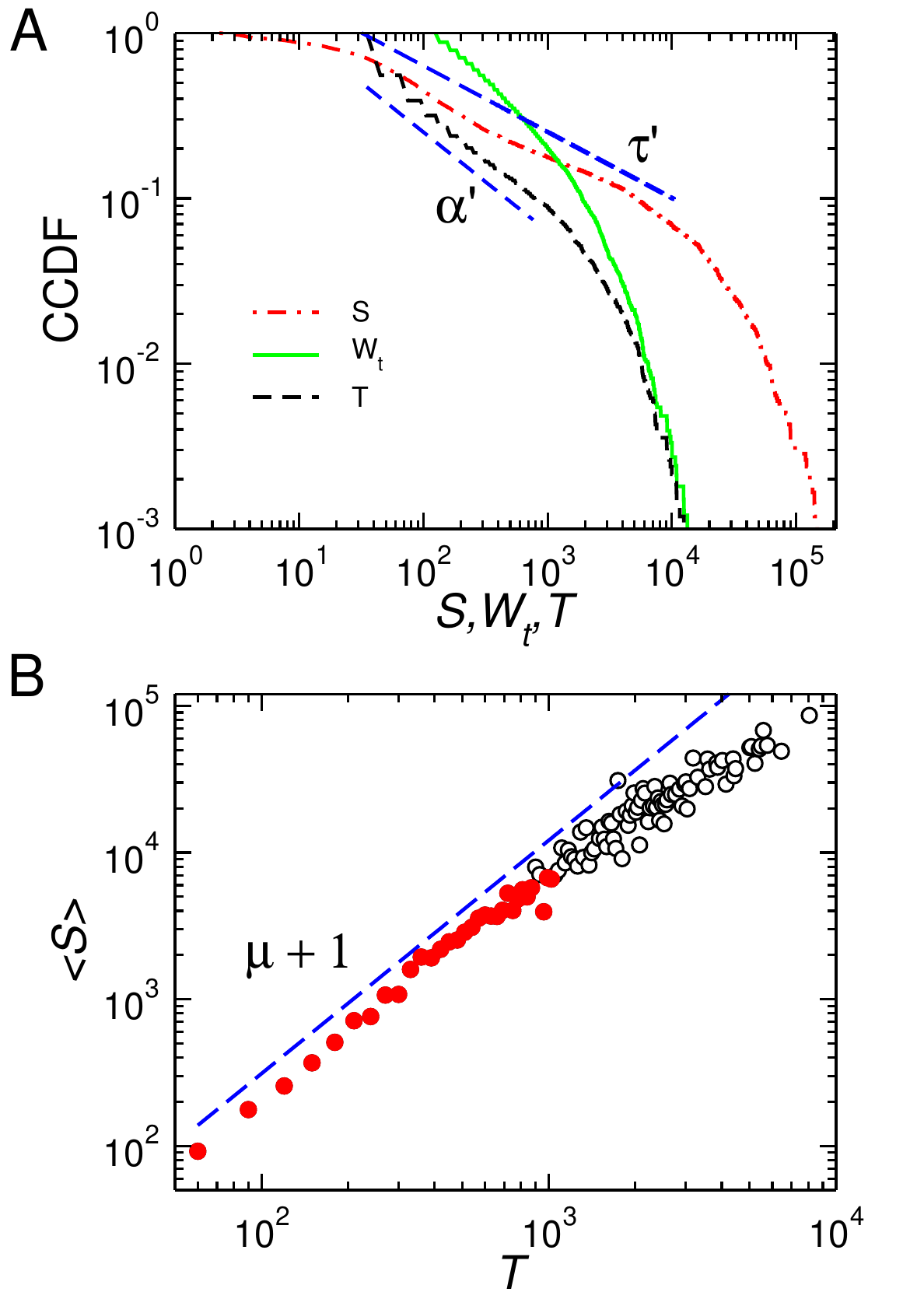} 
\caption{Scaling of activity events in a single subject (same dataset as in Fig. 1). Panel A: The complementary cumulative distribution function (CCDF) for  event durations (T) and sizes (S) obeys power-laws with exponents $\alpha' =0.70$ and $\tau' =0.44$, respectively (dashed lines). Note  that here the densities are cumulative, thus the  exponents of the respective PDFs are $\alpha =\alpha ' +1 $ and  $\tau =\tau ' +1 $. The waiting time between events falls exponentially. Panel B: The average size of a given duration is well described (for small T) by $\langle S \rangle (T)\sim T^ {\mu+1}$  with $\mu+1=1.59$ (blue dashed line) comparable with results obtained from fitting within the scaling region (red filled symbols) giving $\mu+1=1.61$.}
\end{figure}
 %%%%%%%%%%%%%%%%%%

\section{Results}

For ease of presentation, we will use recordings from a single subject to describe the main results. Nevertheless, results are robust as well as similar for  the entire group of subjects in the study. A typical recording is presented in Fig. 1.  Panel  A shows a full month of continuously recorded activity from this subject, who is particularly regular in her daily routines. The subject wakes up with the alarm clock at 6:45 a.m. on week days and has lunch followed by a short nap each day (between 2:00 p.m. and 4:00 p.m. Panel B displays the time series of the successive increments of the signal $x(n)$, defined as $I(n) = 
x(n+1) - x(n)$. 

The large-scale statistical features of the time series presented in Fig. 1 are already well known. The density distribution of the successive increments $i(n)$ is non-Gaussian, as can be appreciated by a joint plot with a Gaussian distribution of the same variance (Fig. 1, Panel C). It is known that the power spectrum of the activity decays  as $S(f) \sim f^ {\beta}$ \cite{nakamura1,nakamura2}. Because this type of processes are likely to be non-stationary, it is best to estimate the exponents of the spectral density by doing the calculations over the time series of successive increments, whose density distribution is stationary.  For instance, for Brownian motion (which is summed white noise), the power spectrum decays $S(f) \sim f^ {\beta}$ with $\beta= -2$ and for white noise $\beta=0$; the summed time series has an exponent $+2$ larger than the non-summed time series.  As discussed in \cite{malamud}, this can be generalized for all self-affine processes: summing a self-affine time series shifts the theoretical power-spectral density exponent by $+2$, and the reverse process is also true: the differences in consecutive values (the ``first differences'') of a Brownian motion result in white noise, thus taking the first differences shifts the theoretical power-spectral density exponent $\beta$ by $-2$.  In our case, the exponent obtained for the time series of successive increments  $I(n)$ was $\gamma = 0.9$.  Thus, the exponent of the raw data is $\beta = \gamma -2 = -1.1$ \cite{malamud}. For comparison, the spectral densities of the actual signal and of a surrogate obtained after randomly shuffling the increments are jointly displayed in Panel D of Fig. 1.  

To further study the time series from the perspective of individual bursts of activity, we introduce the definition of an event. We consider the time series of activity $x(n)$ and select a threshold value $U$  to be vanishingly small. An event is defined by the consecutive points starting when $x(n) > U$ and ending when $x(n) < U$. This is equivalent to the definition of avalanches in other contexts \cite{bak,laurson}.  In the following part, we will be concerned primarily  with the statistics of event lifetimes $T$, as well as of their average size $S$ and shape. 
%%%%%%%%%%%%Figure 3%%%%%%%%%%%%%%%
\begin{figure}[th]
\centering
\includegraphics[width=.41 \textwidth]{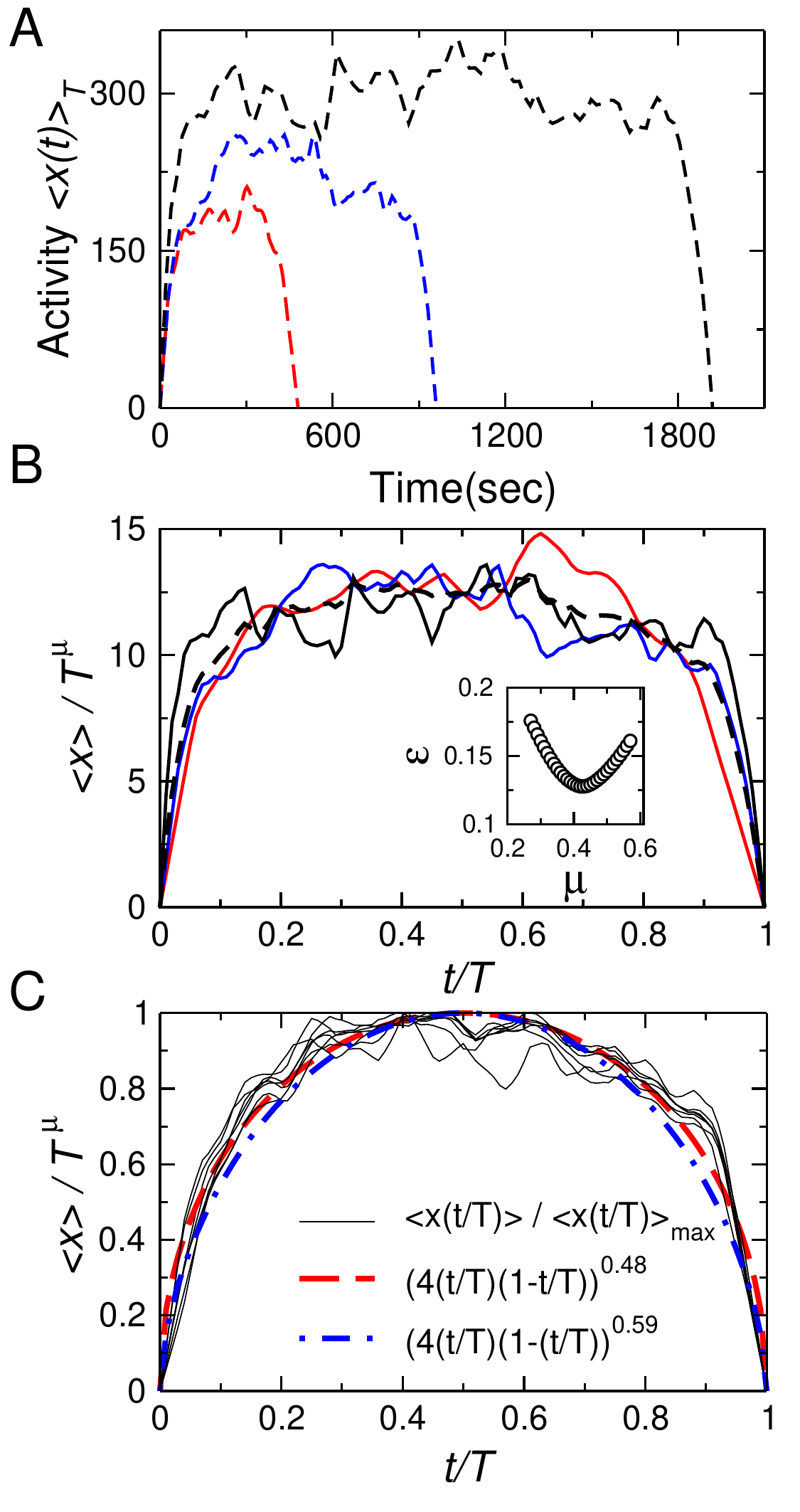} 
\caption{Collapse of events of different duration into a single functional form. Panel A: Three examples of typical events of duration T=480, 960 and 1920 sec.. Panel B: The heterogeneous events shown in Panel A can be collapsed onto the average shape (dashed black line) by normalizing $t$ to $t/T$ and $\langle x(t) \rangle$ to $\langle x(t) \rangle /T^{\mu}$. The inset shows the cumulative variance for a range of $\mu$. Panel C: The average event shape, i.e., $f_{shape}(t/T)$, recovered from six data sets (thin lines). The best fit using an inverted parabola is shown as a red dashed line ($\mu=0.49$) as well as the one expected from the critical exponent $\mu=0.59$ as a dot-dashed blue line. } 
 \end{figure}
%%%%%%%%%%%%%%%%%%%%%%%%%%%%%%%%
In all subjects, we found that the distributions of event durations and sizes (defined by the area, i.e., the integral of the signal corresponding to the individual events) can be well described, for relatively small values, by a power-law (Fig. 2, Panel A). In contrast, the distribution of waiting times between events demonstrated an exponential decay. In addition to the scale invariance, we  found that  the longer an event lasted,  the stronger the motor activity executed by the subject. The plot of average event size $\langle S \rangle$ as a function of duration $T$  follows a power-law (for small values of $T$) described by $\langle S \rangle (T) = T^{\mu +1}$ with $\mu +1 = 1.59$.  The exponents in this power-law are robust across subjects and  to changes of threshold over a reasonable range of values. 

This type of scaling is well known in the statistical mechanics of critical phenomena \cite{bak}.  Examples range from earthquakes \cite{gutemberg} to active transport processes in cells \cite{protein}, crackling noise \cite{sethna},  the statistics of Barkhausen noise in permalloy thin films \cite{papaniko} and plastic deformation of metals \cite{metals}. In all these cases, the distributions obey universal functional forms:
\begin{equation}
f(S)\sim S^{-\tau}, 
\end{equation}
\begin{equation}
 f(T) \sim T ^{-\alpha},
 \end{equation}
\begin{equation}
\langle S \rangle (T)\sim T ^{1/{ \sigma \nu z}},
\end{equation}
where $f$ denotes the probability density functions of  the size of the event $S$ and its duration $T$, and $\langle S \rangle (T)$ is the expected size for a given duration. The parameters $\tau$, $\alpha$  and $1/{ \sigma \nu z}$  are the critical exponents of the system and are expected to be independent of the details, being related to each other by the scaling relation:

\begin{equation}
 {{\alpha - 1}\over {\tau-1} }= {{1} \over { \sigma \nu z}}.
\end{equation}

We found that the empirical exponents very closely fulfill the expression above. Using the fitting approach introduced by Clauset \cite{clauset} in the scaling regions depicted in Panel A of Fig. 2,  we found  $\tau =1.44$  and $ \alpha =1.70$ . Thus, from  Eq. (4) a value of $1/{ \sigma \nu z} = \mu +1 = 1.59 $ is expected. The experimental data points are very close to this theoretical expectation (dashed line), especially for the relatively small T values within the scaling region of Panel A (where a linear fit estimates $\mu+1=1.61$), while those for relatively larger T values (corresponding to the cutoff of the distributions) are a bit apart, probably due to undersampling.  After repeating this analysis for all subjects in our sample,  the average exponents were all within $5\%$ of the reported values.

From scaling arguments, it is expected that the average shape of an event of duration $T$ $\langle x (T , t) \rangle$ scales as :
\begin{equation}
\langle x(T,t) \rangle= T ^{\mu} f_{shape}(t/T).
\end{equation}

Thus, the shapes of events of different durations $T$ rescaled by $\mu$  should collapse on a single scaling function given by $f_{shape}(t/T)$.  Note that  $\mu$  corresponds in this context to the wandering exponent (i.e., the mean squared displacement) of the activity \cite{baldassarri,colaiori}.   

%%% Figure 4 %%%%%%%%%%%%%%%%%
\begin{figure}[th]
\centering
\includegraphics[width=.41 \textwidth]{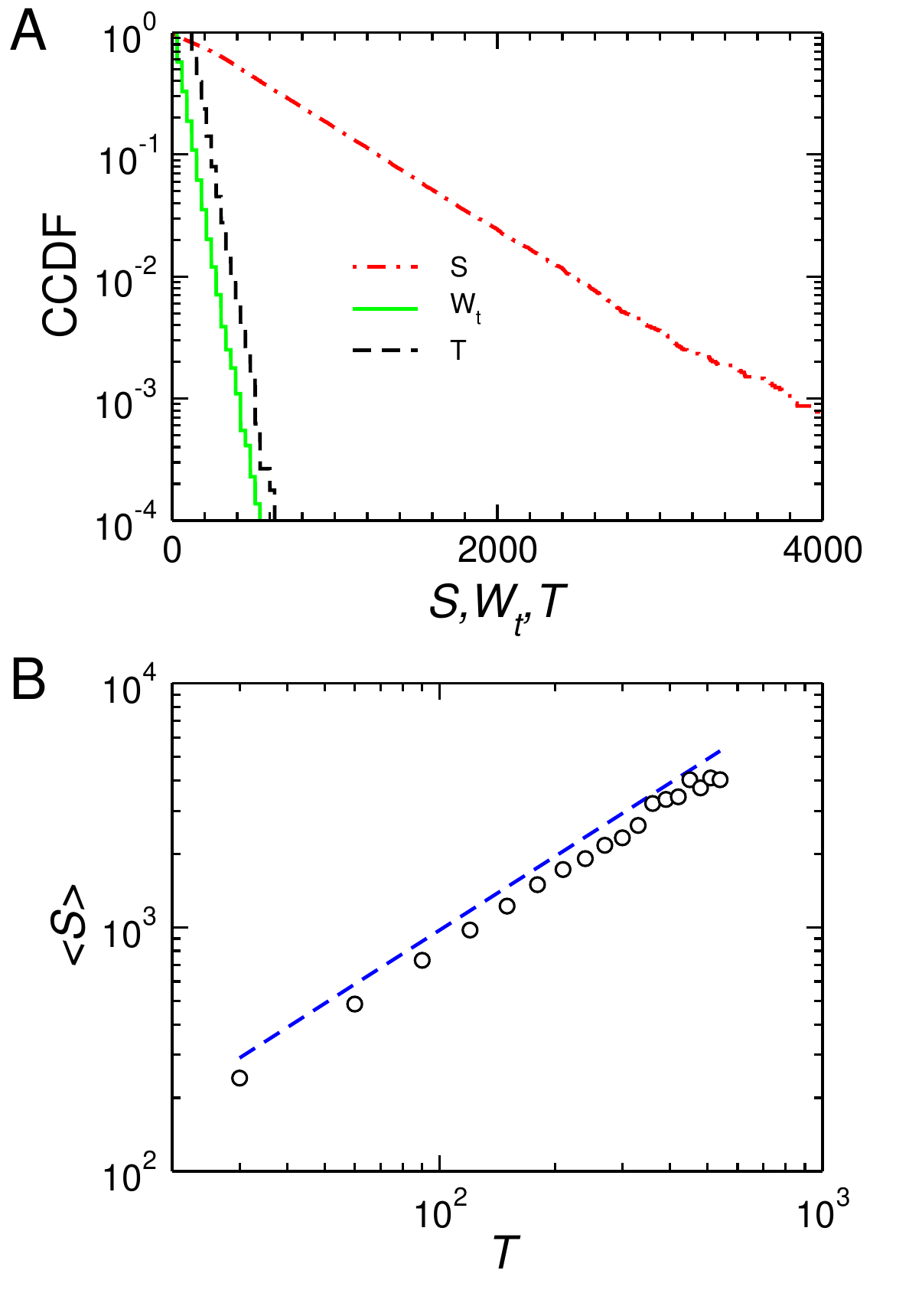} 
\caption{Scaling is absent in a null model resulting from defining events after randomly reordering the time series $x(n)$. Panel A: Density distributions  (CCDF) for event duration, size, and waiting time. All the distributions are exponential (note the logarithmic-linear scale).  Panel B: The expected average size for a given duration in the null model is a linear function of T (the dashed line represents the fit with slope 1), therefore,  $\mu =0$ and there is no collapse.
} 
 \end{figure}
%%%%%%%%%%%%%%%%%%%%%%%%%%
Examples of this collapse are presented in Panels A and B of Fig. 3.  Considering the number of events  here averaged (in the order of  $N\sim10^2$), the data collapse is quite satisfactory, while the value of the exponent ($\mu=0.48$) does not exactly match the one predicted in Eq. (4), $\mu=0.59$ (likely a consequence of insufficient sampling). To determine the generality of our results, we extended this analysis to six other data sets. For each data set, the value of $\mu$ was first determined. Subsequently, the $x(T,t)$ obtained from the events were rescaled with $T^\mu$ and their average computed. To account for individual differences in mean activity, shape functions were  normalized by their mean value. The results for  the  six datasets are presented in Panel C of Fig. 3. They can be accurately described by an inverted parabola, as in other systems previously studied using this method. The best fit disagrees with the empirical functions near their peak, the latter being flatter, likely an effect related to saturation observed in long events.

Finally, we turn to discuss simple null models. We consider two extreme cases, in both of them the raw time series are randomly shuffled to remove serial correlations. In the first case, we remove all temporal correlations by randomly reordering  $x(n)$, thus attaining a flat power spectral density.  After repeating the above analysis in this surrogate data set, it becomes clear (as shown in Fig. 4) that the scale invariance is absent in all the statistics under study: size $S$, waiting time $W_t$ and duration $T$ of events (note that the distributions are here plotted using a logarithmic-linear scale). Results in Panel B show that  $\mu + 1=1$, thus $\mu =0$, implying that there is not collapse, because with $T ^{\mu}=1$ in Eq. (5), the amplitude of the individual events remains invariable. To consider the second case,  we need first to reorder randomly the time series of increments  $I(n)$ and then proceed to integrate the increments. Since each increment is now a random variable, the power spectral density for this surrogate process obeys $f{^{\beta}}$ with $\beta=-2$ , and as shown analytically by Baldassari et al. \cite{baldassarri}, for this case $\mu=1/2$ and the scaling function is a semicircle. Please note that the fluctuations of human activity described here differ from a simple auto-regressive process: indeed successive increments $I(n)$ are anti-correlated and the power spectral density corresponds to non-trivial power law correlations  (i.e., $\beta \neq -2$).

\section{Discussion}

The present findings can be summarized by six stylized facts describing bursts of human activity: I) the spectral density of the time series of activity $x(t)$  obeys a power law, with exponent  $\beta \sim 1$;  II) successive increments $I(n)$ are anti-correlated with a spectral density obeying a power law with exponent $\gamma \sim 1 $, which corresponds to a spectral density for the raw data $f{^{\beta}}$ with $\beta \sim -1$; III) the PDF of the increments $I(n)$ is definitely non-gaussian; IV)  the PDF of  duration and sizes of events obeys truncated power laws with exponents $1<\tau<2$ and $1<\alpha<2$; V) the average size of the events scales with its lifetime $T$ as $ \langle S \rangle (T)\sim T ^{\mu}$, where $\mu +1=(\alpha-1)/(\tau-1)$; VI) the time series of individual events can be appropriately rescaled via a transformation of its duration $T$  and amplitude $x(t)$ onto a unique functional shape: $\langle x(T,t) \rangle= T ^{\mu} f_{shape}(t/T)$.

We are aware that these observations are novel only for human activity, because similar statistical regularities of avalanching activity are well known for a large variety of inanimate systems \cite{laurson, papaniko, sethna,neuro1}. The rescaling of the average shape is not surprising because, placed in the appropriate context, it can be traced back to Mandelbrot's study of the fractal properties of self-affine functions \cite{Mandelbrot}. A curve or a time series are said to be self-affine if a transformation can be found,  such that rescaling their $x ,  y$ coordinates by $k$ and $k^\mu$, respectively, and the variance in $y$ is preserved (with $\mu=1$ corresponding to self-similarity). In that sense, the successful collapse of the events shape is a trivial consequence of the overall self-affinity of the $x(t)$ time series.

Thus, it is clear that the existence of the scaling uncovered here is not informative per se of the type of mechanism behind: scale-invariance can be constructed via different processes, ranging from critical phenomena \cite{bak} to simple stochastic auto-regressive dynamics \cite{baldassarri,colaiori}. What is  then the mechanism by which the above six facts are generated? 
 
It seems that this question cannot be easily answered by the type of experiments reported here. Fluctuations of this type could have either an intrinsic (i.e., brain-born) origin but also could be the reflection of a collective phenomena (including humans and its environment).  In either case, the correlations observed seem to reject the  case of independent random events starting and stopping human actions, because neither the distribution of the increments $I(n)$, nor the exponents match the case of a random walk. In terms of brain-born process, it is hard to accept some of the implications of the scaling function in the activity shape.  The average parabolic shape means that the very beginning of the motion activity contains information about how long the activity will last, in the same sense that the initial trajectory of a projectile predicts when and where it will land. This proposal is hardly realistic, because there is hardly a reasonable physiological argument in  support of any  motor planning for  the length of time we are observing ($\sim10^3$ secs). In terms of collective processes, the results here suggest that the interaction with other humans could determine when and where, on the average, we start and stop moving. 

Despite our current relative ignorance, a possibility that sounds interesting is to determine in children, as they grow, if their behavioral product of parental (and otherwise) education are reflected in the shape of their individual scaling function. This seems reasonable given the fact  that ``tireless running around'' is  almost a definition of early age well-being, which gives way to less hectic activity as children mature. In the same line of thoughts, if changes in the scaling function can be quantitatively traced to behavioral changes, one could also consider to explore applications of these techniques to monitor eventual progress in the treatment of hyperactivity disturbances such as in the subjects affected by the Attention Deficit Hyperactivity Disorder syndrome. The converse, i.e., cases in which the average activity diminish, as in elderly subjects shall be also explored. Further experiments and analysis should shed light on these possibilities. In, the meantime, the present results provide a guide and six important constraints for the models that should best capture the physics (and biology) of the process.

\begin{acknowledgements}
Work supported by National Science Center  of Poland (ncn.gov.pl, grant DEC-2011/02/A/ST1/00119);  State Secretary for Research and Development (grants PSI2010-19372 and PSI2013-48260)  from Spain and by CONICET from Argentina.)
\end{acknowledgements}

\end{document}